\def\msun{{\,M_\odot}}
\def\lsun{L_{\odot}}

\def\etal{et al.$\,$}
\def\deg{^\circ}
\def\vtsn{\vartheta _{\rm SN}}
\def\ros{\rho _*}

\def\mas{M _*}
\def\lb{L_{\rm B}}
\def\lx{L_{\rm X}}
\def\lgm{L_{\rm g}}

\def\ls{L_{\sigma}}
\def\lr{L_{\rm rot}}
\def\lsn{L_{\rm SN}}
\def\zb{\zeta}
\def\phh{\Phi_{\rm h}}
\def\mah{M_{\rm h}}
\def\phit{\Phi_{\rm tot}}
\def\vphi{\overline {v_\varphi}}
\def\vdphi{\overline {v^2_\varphi}}
\def\sphi{\sigma_\varphi}
\def\sr{\sigma_{\rm R}}
\def\sz{\sigma_{\rm z}}
\def\ugv{{\bf v}}
\def\usv{{\bf v_*}}
\def\mgv{{\bf m}}
\def\alst{\alpha_*}
\def\alsn{\alpha_{\rm SN}}
\def\SX{\Sigma_{\rm X}}
\def\SS{\Sigma_*}
\def\ah{a_{\rm h}}
\def\bh{b_{\rm h}}

\def\tfil{t_{\rm fil}}
\def\tcc{t_{\rm cc}}
\def\vrot{v_{\rm rot}}
\def\sigc{\sigma_{\rm c}}
\def\elrho{{\cal E}_{\rho}}
\def\mdisk{M_{\rm disk}}
\catcode`\@=11
\def\gsim{\ifmmode{\mathrel{\mathpalette\@versim>}}
    \else{$\mathrel{\mathpalette\@versim>}$}\fi}
\def\lsim{\ifmmode{\mathrel{\mathpalette\@versim<}}
    \else{$\mathrel{\mathpalette\@versim<}$}\fi}
\def\@versim#1#2{\lower 2.9truept \vbox{\baselineskip 0pt \lineskip
    0.5truept \ialign{$\m@th#1\hfil##\hfil$\crcr#2\crcr\sim\crcr}}}
\catcode`\@=12
\def\spose#1{\hbox to 0pt{#1\hss}}
\def\lta{\mathrel{\spose{\lower 3pt\hbox{$\mathchar"218$}}
     \raise 2.0pt\hbox{$\mathchar"13C$}}}
\def\gta{\mathrel{\spose{\lower 3pt\hbox{$\mathchar"218$}}
     \raise 2.0pt\hbox{$\mathchar"13E$}}}

\documentstyle[12pt,aasms4]{article}

\slugcomment{In press on ApJ (Main Journal)}

\lefthead{D'Ercole and Ciotti}
\righthead{Gas flows in S0 galaxies}

\begin{document}

\title{Decoupled and Inhomogeneous Gas Flows in S0 Galaxies}

\author{A. D'Ercole and L. Ciotti}
\affil{Osservatorio Astronomico di Bologna, via Zamboni 33, 40126 Bologna, 
      ITALY}



\begin{abstract}

A recent analysis of the {\it Einstein} sample of early-type galaxies has
revealed that at any fixed optical luminosity $\lb$ S0 galaxies have
lower mean X-ray luminosity $\lx$ per unit $\lb$ than ellipticals.
Following a previous analytical investigation of this problem (Ciotti
\& Pellegrini 1996), we have performed 2D numerical simulations of the
gas flows inside S0 galaxies in order to ascertain the effectiveness
of rotation and/or galaxy flattening in reducing the $\lx/\lb$ ratio.

The flow in models without Supernova (SNIa) heating is considerably
ordered, and essentially all the gas lost by the stars is cooled and
accumulated in the galaxy center. If rotation is present, the cold
material settles in a disk on the galactic equatorial plane.  Models
with a time decreasing SNIa heating host gas flows that can be much
more complex.  After an initial wind phase, gas flows in energetically
strongly bound galaxies tend to reverse to inflows. This occurs in the
polar regions, while the disk is still in the outflow phase. In this
phase of strong decoupling, cold filaments are created at the
interface between inflowing and outflowing gas.  Models with more
realistic values of the dynamical quantities are preferentially found
in the wind phase with respect to their spherical counterparts of equal
$\lb$. The resulting $\lx$ of this class of models is lower than
in spherical models with the same $\lb$ and SNIa heating. At variance
with cooling flow models, rotation is shown to have only a marginal
effect in this reduction, while the flattening is one of the driving
parameters for such underluminosity, in accordance with the analytical
investigation.

\end{abstract}




%

\section{Introduction}

It is well known that normal early-type galaxies are X-ray emitters,
with 0.2--4 keV luminosities ranging from $\sim 10^{40}$ to $\sim
10^{43}$ erg s$^{-1}$. The X-ray luminosity $\lx$ is found to
correlate with the blue luminosity $\lb$ ($\lx\propto \lb^{2.0 \pm
0.2}$), but there is a large scatter of roughly two orders of
magnitude in $\lx$ at any fixed $\lb > 3\times 10^{10}\lsun$
(\cite{fab89}; Fabbiano, Kim, \& Trinchieri 1992).

The trend and the scatter in the $\lx -\lb$ diagram have been
successfully reproduced by the time-dependent, 1D hydrodynamical
models of Ciotti \etal (1991, hereafter CDPR). In these models the gas
lost by the evolving stars is heated to X-ray temperatures by stellar
random motions and SNIa explosions.  Given the assumed time dependence
of both mass loss rate ($\propto t^{-1.36}$) and SNIa rate ($\propto
t^{-1.5}$), the CDPR models can evolve through up to three different dynamical
phases, i.e., from winds to subsonic outflows to inflows (in the
following we call this kind of evolution the WOI sequence).  CDPR
introduced the {\it global} parameter $\chi$ in order to predict the
dynamical status of the gas flows in non-rotating, spherical galaxies.
This parameter was defined as the ratio between the power required to
extract steadily the gas supplied by the mass loss of the evolving
stellar population from the galaxy potential well, $\lgm$, and the
power supplied to the gas by the thermalization of the stellar random
motions and by the supernova heating ($\ls$ and $\lsn$,
respectively). Due to the temporal evolution of the various source
terms, $\chi$ is an increasing function of time. Numerical simulations
showed the correspondence between $\chi <1$ and the presence of a wind
phase (i.e., of a low $\lx$), and the onset of a high $\lx$ soon after
$\chi$ becomes $>1$.
 
A recent analysis of the complete $Einstein$ sample of early-type
galaxies showed that on average S0 galaxies have lower $\lx$ and
$\lx/\lb$ than ellipticals (Es) (Eskridge, Fabbiano, \& Kim 1995a,b).
A possible explanation of this fact could be that S0s and
non-spherical Es are less able to retain hot gaseous haloes than
rounder systems of the same $\lb$.  This could be due to a lower
gravitational energy, because of their higher rotation rate which
decreases the effective potential, or their different mass
distribution. A theoretical analysis of this problem -- based on the
global energetics of the X-ray emitting gas -- was carried out by
Ciotti \& Pellegrini (1996, hereafter CP). They generalized the global
parameter $\chi$ for flows in flat and rotating two-component galaxy
models. In particular, they added to $\chi$ the contribution of the
ordered rotation of the galaxy stellar component, $\lr$, showing that,
at least from a global point of view, rotation alone cannot change an
energetically bound gas flow to an unbound one (see \S 2.3).  On the
contrary, using two-component Miyamoto-Nagai models (1975, hereafter
MN), and under the assumption of an identical rate of SNIa per unit
$\lb$ in S0s and Es, CP showed that flat galaxies can have a much
lower $\chi$ than rounder ones of the same luminosity, due essentially
to their flattening, even in the presence of massive spherical dark
haloes. More precisely, they showed how a realistic flattening can
produce a decrease of $\lgm$ of $\sim 20\%$, a result which can be
sufficient to change an inflow into a wind. So, the effect of the
flattening was suggested as a possible explanation for the S0
underluminosity.  Incidentally, note that the decrease in the
gravitational energy that is released by a stationary inflow at the
bottom of the galactic potential well, associated to the decrease of
$\lgm$, does not lower significantly the $\lx$ of pure cooling flow
models (cfr. Fig.1 in CDPR).

Despite the possible agreement between the observations and the
results of CP, this purely energetical approach can be seriously
questioned considering the complexity of 2D gas flows and the fact
that $\chi$ is a global quantity. In fact {\it decoupled flows} -- in
which at the same time different parts of the galaxy are in inflow and
outflow -- may in principle develop in the CDPR scenario in such a way
that $\lx$ remains high even if $\chi <1$; if this is the case, $\chi
<1$ cannot be used as a reliable indicator for a low X-ray luminosity,
and the analysis of CP is seriously questioned.

2D numerical simulations of gas flows in early-type galaxies were
actually carried out by Kley \& Mathews (1995, hereafter KM) and
Brighenti \& Mathews (1996, hereafter BM).  These simulations, however,
considered only pure cooling flows for which the flow decoupling is
not expected.  KM considered rotating spherical King models plus a
spherical quasi-isothermal dark halo, without SNIa heating.
Successively BM considered oblate galaxies with different axial
ratios, sustained by various amounts of ordered and disordered kinetic
energies. They assumed a SNIa rate $\propto t^{-1}$, with a present
day value of one-tenth of that proposed by Tammann (1982). Their
models show that the gas flow and the associated $\lx$ are only
slightly dependent on the flattening of the galaxy (in agreement with
the CP analysis applied to the {\it inflow} case), and that $\lx$ is
reduced by increasing the amount of ordered rotation.  These 2D
hydrodynamical models do not solve problems such as the disposal of
excessive amounts of cold gas (a problem already encountered in 1D
pure cooling flow models) on an equatorial disk, which is actually
unobserved, and the excessively flat shape of the X-ray isophotes when
compared with the observations.

Motivated by the given arguments, we adopt the same input physics
used for the hydrodynamical equations in CDPR, in order to explore
whether the WOI sequence is still achieved in flat galaxies, how such
a sequence is influenced by galaxy flattening and rotation, how
important the decoupling in the flow is, which observational
consequences are predicted in this case, and finally whether the
energetical explanation of CP is exhaustive.  This is accomplished with
two different classes of models. In the first we compare the behavior of
the gas flow in spherical and flat galaxies that are {\it
energetically globally equivalent} (both with and without SNIa
heating).  In the second class of simulations we investigate the gas
flows in realistic galaxy models of S0s, also in order to explore the
possibility of flow decoupling in galaxies with $\chi <1$ and its
effects on $\lx$.

We thus study the evolution of gas flows in the galaxy models
constructed and described in detail by CP, i.e., axisymmetric stellar
and dark mass distributions with the MN shape, and in which the
internal dynamics is determined by the solution of the associated
Jeans equations, allowing for variable amounts of ordered rotation and
velocity dispersion.

In Section 2 we briefly describe the models used in the simulations.
In Section 3 the results obtained for models with different amounts of
ordered rotation, SN heating, flattening, and characteristic
scale-length are presented. A discussion, considering also some recent
observational results is given in Section 4, and in Section 5 the main
conclusions are summarized.

\section{The models}

\subsection{The galaxy density distribution}


The two-component mass models used in our simulations and their
internal dynamics are fully discussed in CP, which includes the explicit
analytical formulae for the various terms in the virial theorem
and for $\chi$ are given. Here only the basic formulae are reported.  The
stellar density distribution is given by 
\begin{equation} 
\ros=\left({\mas b^2\over 4\pi} \right) {aR^2+(a+3\zb)(a+\zb)^2\over
\zb^3[R^2+(a+\zb)^2]^{5/2}}, 
\end{equation} 
and its potential is
\begin{equation} 
\Phi_* =-{G\mas\over\sqrt{R^2+(a+\zb)^2}},
\end{equation} 
where $\zb=\sqrt{z^2+b^2}$, and $(R,\varphi,z)$ are the cylindrical
coordinates (MN)\footnote{When $a=0$ the density distribution (1)
reduces to the Plummer (1911) model, and $b$ becomes its effective
radius. So, $b$ can be roughly considered a characteristic
scale-length, while the galaxy flattening is determined by the
dimensionless ratio $s=a/b$.}.  The dark matter halo is described by
another MN density distribution characterized by $\mah$, $\ah$ and
$\bh$ The associated Jeans equations are solved in the total potential
$\phit=\Phi_*+\phh$, and under the assumption that the distribution
function depends only on the specific energy and the angular momentum
component along the $z$-axis (see, e.g., Binney \& Tremaine 1987,
p.197).  With this assumption the radial and axial velocity
dispersions are equal, i.e., $\sr=\sz\equiv\sigma$, and the only
non-zero streaming motion can be in the azimuthal direction
($\varphi$).  The intrinsic degeneracy in the azimuthal velocity
component is removed adopting the Satoh (1980) $k$-decomposition, i.e.,
\begin{equation}
\vphi^2=k^2(\vdphi-\sigma^2),
\end{equation}
and 
\begin{equation}
\sphi^2\equiv\vdphi-\vphi^2=k^2\sigma^2+(1-k^2)\vdphi,
\end{equation}
with $0\leq k\leq 1$, where $\vphi$ is the modulus of the ordered
rotational velocity of the galactic stellar component. In the case
$k=0$ no net rotation is present and all the flattening is due to
$\sphi^2$; $k=1$ corresponds to the isotropic rotator.  The analytical
solutions of the Jeans equations can be obtained for $\bh = b$. In
simple case of $\ah=a$ the formulae for the velocity components are
expressed in terms of elementary functions and are given in CP
together with their edge-on projection.

\subsection{The hydrodynamical equations}

The equations describing the gas flows in our simulations are
\begin{equation}
{\partial\rho\over\partial t}+\nabla \cdot (\rho\ugv)=\alpha\ros, 
\end{equation}
\begin{equation}
{\partial\mgv\over\partial t}+\nabla \cdot (\mgv\otimes\ugv)=
-(\gamma-1)\nabla E+{\bf g}\rho+\alpha\ros\usv
\end{equation}
\begin{equation}
{\partial E\over \partial t}+\nabla \cdot (E\ugv)=
-(\gamma-1)\,E\nabla\cdot\ugv-L+\alpha\ros(\epsilon_{\circ} 
+{1\over 2}\Vert\ugv-\usv\Vert^2),
\end{equation}
where $\ugv$ is the fluid velocity, and $\rho$, $\mgv$, and $E$ are
the mass, the momentum, and the internal energy of the gas per unit
volume, respectively.  The source terms on the r.h.s. of equations
(5)-(7) describe the injection of mass, momentum and energy in the gas
due to the mass return and energy input from the stars.  $\alpha
(t)=\alst(t)+\alsn(t)$ is the specific mass return rate from stars and
SNIa respectively, with $\alst\propto t^{-1.36}$ and $\alsn\propto
t^{-1.5}$; $\epsilon_{\circ}$ is the injection energy per unit mass
due to the stellar random motions and to SNIa explosions. $\usv$ is
the bulk velocity of the stars, whose modulus is given by equation
(3), ${\bf g}=-\nabla\phit $ is the gravitational acceleration due to
the total mass distribution, and $\gamma =5/3$.  Finally, $L=n_{\rm
e}n_{\rm p}\Lambda (T)$ is the cooling rate per unit volume, where for
the cooling law $\Lambda (T)$ we follow the prescription of Mathews \&
Bregman (1978).  For more details on the exact form of $\alst(t)$,
$\alsn(t)$ and $\epsilon_{\circ}$, see CDPR and KM.

To integrate the set of equations we used a second-order, upwind, 2D
extension of the numerical code described in CDPR, coupled with a
staggered, spherical, Eulerian grid. Given the symmetry of the
problem, the grid covers polar angles $0\deg \leq \theta \leq
90\deg$ with 40 equally-spaced angular zones, and reflecting boundary
conditions are applied at $\theta=0\deg$ and $\theta=90\deg$. The
radial coordinate covers the range $0\leq r =\sqrt{R^2+z^2}\leq 100$
kpc with 80 logarithmically spaced zones, and a free outflow from the
grid is allowed at the outer edge. At the inner edge reflecting
boundary conditions are set.  We assume the model galaxies to be
initially devoid of gas due to the previous activity of the Type II
SNe. A discussion on the reliability and implications of this
assumption is given in CDPR.

\subsection{The $\chi$ parameter}

The main ingredients of the global parameter $\chi$ are
presented here.  The SNIa heating evolution is parameterized as
$\lsn=10^{51}R_{\rm SN}\lb$ erg s$^{-1}$, with
\begin{equation}
R_{\rm SN}=0.88 h^2\,{\rm SNU}\,\vtsn 
{\left (t\over 15 {\rm Gyr} \right )}^{-1.5},
\end{equation}
where $h=H_{\circ}/100\,{\rm km}\,{\rm s}^{-1}\,{\rm Mpc}^{-1}$, and 1
SNU = 1 SN per century per $10^{10}\lsun$ (cfr. CDPR).  When $t=15$
Gyr, $\lb=10^{10}\lsun$ and $\vtsn=1$, the standard SNIa rate is
recovered (Tammann 1982)\footnote{Note that in our models the assumed 
$\lb$ is independent of $h$. As a consequence $\lsn$ $depends$ on $h$. 
In the paper we adopt $h=1/2$}.

The power $\lgm$ required to extract steadily from the galaxy the mass
losses from the aging stellar population is
\begin{equation}
\lgm =\alpha(t)\int \ros |\phit |{\rm d}V.
\end{equation} 
The thermalization of the stellar random motions heats the gas at a power 
$\ls$ given by 
\begin{equation}
\ls={\alpha(t)\over 2}\int \ros (2\sigma^2+\sphi^2){\rm d}V,
\end{equation}
while the power $\lr$ related to the ordered fraction of the kinetic
energy is given by
\begin{equation}
\lr={\alpha(t)\over 2}\int \ros\vphi^2 {\rm d}V.
\end{equation}

Unfortunately this last term cannot be inserted in the $\chi$
definition in a simple way.  In fact, two opposite and extreme
scenarios can be assumed: one in which also $\lr$ is thermalized
(contributing to the gas heating together with $\ls$ and $\lsn$), and
another in which the only effect of $\lr$ is to reduce the depth of
the effective galactic potential well.  As a consequence, the global
parameter $\chi$ is defined as
\begin{equation}
\chi={\lgm -\gamma\lr\over\lsn +\ls+(1-\gamma)\lr},
\end{equation}
with $0\leq\gamma\leq 1$. Note that if $\gamma=0$, i.e., in the case
of {\it complete thermalization}, the virial theorem implies that $\ls+\lr$
depends only on the galaxy structure and $\chi$ is independent of the
amount of ordered rotation (in our case of the value of the parameter
$k$).  When instead $\gamma =1$, i.e., in the case of {\it cold
rotation}, the effect of rotation is maximum; CP showed that also in
this case, for realistic galaxy models the differences in the value of
$\chi$, for a null and a maximum ordered velocity configuration, are
small, usually less than $10\%$.  Moreover whenever $\gamma>0$ the
introduction of rotation decreases a $\chi <1$ and increases a $\chi
>1$. In conclusion, from the CP analysis it results that rotation
alone cannot invert the global energy balance of a gas flow, but can
help the occurrence of the wind when $\chi<1$.  Due to the small
dependence of $\chi$ on $\gamma$, through the paper we assume $\gamma
=0$.


\section{The results}

\subsection{Model A1: $\vtsn=0.3$ and $k=1$}

We describe here in detail the main characteristics of the flow hosted
by a galaxy model in which ordered rotation ($k=1$) and SNIa heating
in accordance with the latest estimates of Cappellaro et al. (1997,
$\vtsn=0.3$) are present. We assume $\mas=2.75\times 10^{11}\,\msun$,
$\mas/\lb =5.5$, so that $\lb=5\times 10^{10}\,\lsun$, and $a=b=3$
kpc.  The dark halo has the same profile of the stellar distribution,
i.e., $\ah=a$ and $\bh=b$, and $\mah/\mas=2$ (see Table 1). The
three-dimensional central velocity dispersion is $\sigc=330\,{\rm
km}\,{\rm s}^{-1}$, and the maximum of the ordered rotational velocity
is $\vrot=360\,{\rm km}\, {\rm s}^{-1}$.  This set of parameters is
such that $\chi=1$ at the present epoch, in order to compare as much as
possible the present results with those obtained for the so-called
{\it reference model} discussed in CDPR (hereafter KRM, $\chi _{\rm
KRM}=1.1$). $\lb$ is also the same of the KRM, so that the mass return
rate of the two models is exactly the same.  Note how it turns out
that, in order to keep $\chi=1$, model A1 must have very high stellar
$\sigc$ and $\vrot$. This is not a surprise: in fact, as already
pointed out by CP, in a realistic flat model with a non-negligible
$\vtsn$ the present-epoch $\chi$ is substantially lower than unity. The same
comment applies to models A1-A4.

\subsubsection{Hydrodynamics}

As a consequence of the early high SNIa energy injection, due to the
assumed $\vtsn$ value and $\alsn\propto t^{-1.5}$, at the very
beginning the flow is in the wind supersonic phase, i.e., the radial
velocity $u_r(R,z)>v_{\rm sound}(R,z)$ throughout the galaxy (Fig.1, top
left panel).

As time increases and the energetic input decreases according to
equation (8), the sonic surface moves outward, with $u_r$ decreasing
earlier in the central regions of the galaxy, and the gas density increases.
Finally, due to the rapidly increasing radiative losses (the so called
cooling--catastrophe), at $\tcc\simeq 2$ Gyr the fluid starts to
revert its motion in a conical region containing the polar axis.  We
call this phenomenon {\it flow inversion}, and the subsequent flow
{\it decoupled inflow}.  The most striking feature of this
evolutionary phase is the formation of cold, dense {\it filaments} at
the boundary where two streams of inflowing and outflowing gas
converge, as can be seen in Fig.1 (top right panel).  This filamentary
phase, which in model A1 suddenly develops at $\tfil\simeq\tcc$, is
rather short ($\lsim 2$ Gyr).  Later on, no more filaments are
generated, and the previous ones are accreted at the galactic
center. However, regions of gas colder than X-ray temperature and with
a low density contrast are present for a much longer time.  During this
evolutionary phase two qualitatively different regimes are present: an
inner zone ($r\lsim 15$ kpc) in which the fluid is rather turbulent,
and an outer zone where the velocity field is smooth (Fig.1, bottom
left panel).  The inflow region becomes progressively larger and
larger. At $t=15$ Gyr the gas is still outflowing in the outskirts of
the disk. The flow in the rest of the galaxy is in a turbulent state,
characterized by very subsonic velocities ($||\ugv|| < 5\,10^6\,{\rm
cm}\,{\rm s}^{-1}$, Fig.1, bottom right panel).


In order to understand better the structure of the cold filaments, we
performed a high-resolution simulation (with 300 radial meshes and 140
angular meshes covering the same physical space as the normal
resolution simulations). Because of the severe limitation due to the
azimuthal time-step, we could not extend the simulation over 15
Gyr. We thus remapped onto the finer grid the hydrodynamical variables
computed on the coarse grid at a time immediately before the filament
formation, and we carried on the simulation for the entire lifetime of
the filament. Fig.2 shows the temperature and velocity field at $t=2$ Gyr
Density and temperature vary along the filament. For $r\lsim
20$ kpc, $\rho_{\rm fil} \sim 6.3\,10^{-27}$ g cm$^{-3}$, and the
density contrast between the cold filaments and the surrounding
density is of the order of 30. The mean temperature is $T_{\rm
fil}\simeq 10^4$K (the lowest allowed temperature in our
simulations). At large $r$ the density decreases and the temperature
increases, but never reaches X-ray temperatures. We did not made
convergence tests, so it is plausible that the filaments are denser
and thinner.  We stress that in our 2D simulations these filaments are
in fact surfaces. This fact should be kept in mind in all the
following discussions\footnote{The formation of such filaments seems
to be a rather general feature in decoupled flows such as this. Melia,
Zylstra, \& Fryxell (1991), for instance, find astonishingly similar
structures in their studies on accretion disk coronae around black
holes.}.


The cold gas accreted during the entire lifetime of the galaxy forms
a small disk, whose radius oscillates around a mean value of 3
kpc. The total mass of the cold disk is $1.5\,10^{10}\msun$, while the
hot X-ray gas is $6.3\,10^8\msun$ (Fig.3b, dotted line 1), much less
than the hot gas in the equivalent KRM in CDPR ($\sim
5.4\,10^9\msun$).

\subsubsection{Energetics}

The comparison with the KRM is instructive. Also for this spherically
symmetric model the present day $\chi\simeq 1$, but $\tcc\simeq 10$
Gyr, much later than in model A1.  From Fig.3d we note that the time
evolution of $\lx$ (dotted line 1) is qualitatively similar to that of
spherical models. After an initial decrease during the wind phase,
$\lx$ increases up to a maximum value at $t=2.5$ Gyr. Later on, $\lx$
decreases again, in pace with the decreasing mass return from the
stars.  When $\lx$ reaches its maximum, a non-negligible fraction of gas
is still outflowing from the galaxy, as in the 1D models.  Note that
at the flow inversion the energy budget ($\chi<1$) indicates that the
SNIa energy injection would be able to sustain a global outflow.  This
means that, because of the strong decoupling of the flow, in this
model the global energetic argument (i.e., the $\chi$ value), is not
a useful indicator of the dynamical state of the gas.

A second important difference between spherical models and model
A1 must be stressed: in the latter, $\lx$ is a factor of four
fainter than $\lsn$ (Fig.3d, solid line), at variance with 1D models
in which $\lx$ may be even greater than $\lsn$ in the inflow
phase. The fainter $\lx$ of model A1 is due to its lower content of
hot gas with respect to the KRM (see \S 3.1.1) . This lower fraction
of hot gas is due to the higher efficiency of the cooling near the
equatorial plane of model A1, a natural consequence of a higher gas
density in this region, which in turn is due to the higher local
stellar density (the effect of rotation will be discussed in \S 3.1.4)

We note however that, although the hot gas mass in model A1 is lower
by about a factor ten with respect to the KRM, $\lx$ does not
diminish by the factor one hundred that one could expect as a rule of
thumb (the cooling per unit volume being $\propto\rho ^2$): the much
lower than expected reduction of $\lx$ is thus due to different hot gas
distributions between the two models.


We finally point out that the noisy temporal behavior of $\lx$ is a
consequence of the density and temperature fluctuations associated to
large eddies in the flow.

\subsubsection{Surface Brightness}

The X-ray surface brightness ($\SX$) evolution of model A1 is shown in
Fig.4 at the same epochs of the shots in Fig.1. The optical isophotes
($\SS$) are also superimposed for comparison, and the galaxy is seen
perfectly edge-on.


During the wind phase, the X-ray isophotes are peanut-shaped, with
regular lobes aligned with the galactic polar direction (Fig.4a). The
elongation of the isophotes decreases steadily with time, they become
rounder and rounder, and, in the inner regions, eventually flatter
than the optical isophotes (Fig.4b,c). The isophotes of $\SX$ at the present
epoch are rather flat in the central regions, and become progressively
circular moving outward. This is due to the well known fact that the
emitting gas accommodates in a potential well that is more spherical
at large $r$. The irregularities in the isophotal shapes are clear
evidence of the inner turbulence (Fig.4d).

Note that once the cold filaments appear (Fig.4b,c), $\SX$
becomes very disturbed: we stress however that such an effect is
magnified in our simulations by the imposed geometry, in which the
filaments are in fact surfaces.  We also computed $\SX$ under
different viewing angles. It turns out that at early epochs the X-ray
isophotes appear still elongated along the optical minor axis, even
for values of the viewing angle as low as $45\deg$ ($90\deg$=edge-on).
Instead, at late epochs, $\SS$ and $\SX$ become very similar as soon as
the inclination angle is slightly lower than $90\deg$.

\subsubsection{Model A2: $\vtsn=0.3$ and $k=0$}

We describe here a model very similar to A1, but with
$k=0$, all the galaxy flattening being due to $\sphi^2$ (see Table
1). For this model the value of $\chi$ is independent of any
assumption on the thermalization of ordered motions [see eq. (12) with
$\gamma=0$], and is equal to unity.

From Fig.3b it is apparent how this non-rotating model has a higher total
gas mass (dashed line 2) than the rotating model A1 (dashed line 1).
This is because rotation - if not entirelly thermalized - decreases a
$\chi <1$ (see \S 2.3), and in fact the largest difference between
dashed line 1 and dashed line 2 takes place during the initial wind
phase, when $\chi$ is substantially lower than unity.  Also the hot
gas mass of model A2 (dotted line 2) is higher than that of model A1
(dotted line 1): this is due to the combined effect of rotation,
already discussed, and of the presence of a slightly more massive cold
disk in model A1, a consequence of angular momentum conservation. This
effect is quite small in the models heated by SNIa, but much more
important in the cooling-flow models (a more detailed discussion is
postponed to \S\S 3.2.2-3.2.3).

The main features of the temporal evolution of $\lx$ are similar for
models A1 and A2 (Fig.3d dotted line 2).  We note however that the
$\lx$ of model A2 is slightly higher than that of model A1, in
accordance with its higher content of hot gas.  The flow dynamical
evolution of model A2 is also quite similar to that of model A1. The
transient phase of filamentary instabilities -- although to a minor
extent -- is still present, and develops approximately at the same
time as in model A1. The absence of rotation does not affect the
presence of oscillations in the temporal evolution of $\lx$, thus
indicating that the turbulence is mainly due to a decoupled
inversion of the gas flow, rather than to rotation. Finally, $\SX$
evolves mainly as in model A1, although the X-ray isophotes appear
more circular in this non-rotating case.

\subsubsection{Models A3 and A4: $\vtsn=0.3$ and $k=1$, but different $a/b$'s}

We focus here on the relationship between the galaxy's flattening and
the development of the cold instabilities.  To investigate this point
we run two models similar to model A1, except for the ratio $a/b$ (see
Table 1).  A quantitative measure of the flattening of a MN model in
its central regions is obtained with a series expansion of
eq. (1). Retaining the quadratic terms the resulting isodensity
surfaces are similar ellipsoids whose ellipticity is
\begin{equation}
\elrho =1-\sqrt{s+5\over (s+1)(s^2+4s+5)},
\end{equation}
and $s=a/b$. Thus $s=1$ adopted in models A1 and A2 corresponds to
$\elrho=0.45$. As a consequence, with $s=0.7$ (and $\elrho=0.36$)
model A3 is slightly more spherical than model A1, and its $\chi=1.2$
is higher, as expected.  The result is that the flow inversion occurs
earlier, and the filamentary phase is already completely developed
at $\tfil=1.4$ Gyr. On the contrary model A4, with $s=1.3$ and and
$\elrho=0.52$ is flatter. Its $\chi=0.94$, and the filamentary phase
starts now at $\tfil=3$ Gyr. For a short time this model shows the
contemporary presence of two cold structures similar to that in
Fig. 2. These two models show how small changes in the galaxy
flattening can change the flow evolution in the CDPR scenario.

\subsection{Model B1: $\vtsn=0$ and $k=1$}

A very different flow evolution is obtained by suppressing the SNIa
heating in a model equal to A1: now $\chi=4$ since the
beginning, and the behaviour is that of a pure cooling flow. The
velocity field is very smooth, thus showing that SNIa's play a
fundamental role in maintaining a boiling hot gaseous halo. In this
case we do not obtain the transient filamentary structure, and $\lx$
decreases smoothly.  The present day $\lx$ is more than one order of
magnitude lower than in the previous SNIa heated cases (Fig.3c, dashed
line), because there is considerably less hot gas in the galaxy
(Fig.3a, dashed line). The cooling gas accumulates now on an extended
cold disk of $\simeq 10^{11}\msun$, approximately ten times more
massive than that of model A1.  Such a disk, present also in the
simulations of BM and KM, is not actually observed in real galaxies.
The X-ray isophotes of model B1 closely follow the optical ones.

\subsubsection{Model B2: $\vtsn=0$ and $k=0$}

Model B2 is analogous to model B1, but no ordered rotation is applied.
In this case $\lx$ is higher than in model B1 at every time (Fig.3c,
dotted line), just due to the larger amount of hot gas (Fig.3a, dotted
line).  At first sight it could seem contradictory that the model with
the lower $\lx$ (B1) has cooled a larger amount of gas, i.e., one
could expect a higher $\lx$, at least in the past. However, although
more mass cools per unit time in the rotating model, the lack of
thermalization of the gravitational energy of the gas which stops on
the rotating cold disk instead of falling into the galaxy center
produces a net quenching in the global $\lx$ (as pointed out by
BM). The $\SX$ isophotes in the non-rotating case are more
concentrated towards the center and rounder than in the rotating case,
as found by BM.


It is of interest to compare our cooling flow models with those of
BM. These authors find a dramatic reduction of $\lx$ when introducing
rotation, but at variance with us, the hot gas mass in their rotating
models decreases only slightly compared to the non-rotating
ones. Thus, the lower luminosity in their rotating models is due
essentially to a redistribution of the hot gas mass, rather than to
its scarcity as in ours.  The origin of this different behaviour is in
the higher flattening of the models used here (i.e., in the presence
of the disk in the MN density distribution).  In fact, in our models a
larger fraction of gas is produced at low-$z$ over the galactic
equatorial plane. When the model is rotating this gas falls on the
disk approximately at the same distance from the center where it was
formed, bacause of angular momentum conservation.  So, on one hand it
releases a much lower gravitational energy, and, on the other, the hot
gas distribution is changed only slightly with respect to the
non-rotating case. The accumulation of the gas on the disk
also significantly increases its cooling, and the final result
is a reduction of
the hot gas mass. Note that also in the BM models the amount of cold
gas increases with an increasing ellipticity of the parent galaxy.

\subsection{Models C1 and D1: $\vtsn=0.3$ and $k=1$}

As mentioneded in the Introduction, we also ran models which are
similar in many aspects
to A1, but with larger $b$'s, in order to have realistic
values for the velocity dispersion (see Table 1).  Since $\ros
(0)=\mas (3+s)/4\pi b^3(1+s)^3$, increasing $b$ at fixed $s$ produces
a reduction in the gas density and thus in the cooling, because of the
lower stellar density. At the same time, because
$\Phi_*(0)=G\mas/b(1+s)$, $\lgm$ is reduced too, favoring the escape
of the gas.

Model C1 ($b=4$ kpc, $\sigc=290$ km/s, $\vrot=309$ km/s, $\chi=0.85$)
develops a decoupled inflow at $5.5$ Gyr, as can be seen from the
evolution of $\lx$ (Fig.3d, heavy dotted line 3). The most striking
feature is again the formation of cold filaments (Fig.5). Surprisingly
enough, these filaments do not form at the moment of flow inversion
($\tcc\simeq 5.5$ Gyr), but quite later ($\tfil\simeq 10.5$ Gyr). At
variance with model A1, they form close to the galactic equatorial
plane (cfr. Fig.2).  Once on the plane, they slowly drift towards the
center and accrete on a small cold disk ($r\sim 4$ kpc, $\mdisk\simeq
5\;10^9\,\msun$). This dynamical phase is characterized by a
conspicuous and steady decrease of $\lx$, a consequence of the
decrease of the hot gas mass (Fig.3b, heavy dotted line 3). While this
effect is certainly real, it is magnified in our simulations for two
reasons.  First, as already pointed out, the filaments are in fact
cylindrical surfaces; second, as usual in 2D simulations, reflecting
boundary conditions on the disk introduce non-physical effects
favoring cold material accumulation.  The $\SX$ evolution is similar
to that of model A1, but postponed in time.

For model D1 ($b=5$ kpc, $\sigc=258$ km/s, $\vrot=277$ km/s,
$\chi=0.7$), the resulting flow is a pure wind all over the galaxy,
lasting for all the Hubble time without any decoupling. All the gas is
hot, and its mass -- as is common in the wind phase -- decreases steadily
and is very low (Fig.3b, heavy line 4).  $\lx$ is also very low and
decreases proportionally to the gas mass (Fig.3d, heavy dotted line
4).  $\SX$ is similar to that shown in Fig. 4a. This run clearly shows
how in a realistic flat model, even with a low $\vtsn$, the flow can
be in the wind phase during its entire evolution.

\subsubsection{Models C2 and D2: $\vtsn=0.3$ and $k=1$, but different $a/b$'s}

We now briefly describe the behaviour of two models similar to
C1 and D1, but with different flattening (see Table 1). The first (C2,
$s=1.3$, $\elrho= 0.52$, $\chi=0.75$) is flatter than model C1, and
consequently has a lower $\chi$. The resulting flow is a wind during all
the time. This model (with $b=4$ kpc) is energetically very similar to
D1 ($s=1$ and $b=5$ kpc), as testified by their $\chi$ values,
and its flow evolution is also similar.

Model D2 ($b=5$ kpc, $s=0.6$, $\elrho=0.33$, $\chi=0.86$) is obtained
from D1 by decreasing its flattening. As a consequence the gas is
more bound as testified by a $\chi$ similar to that of model C1 ($s=1$
and $b=4$ kpc). The resulting evolution is similar to that of model
C1, with the flow inversion and the following development of a
decoupled inflow around $\tcc\simeq 5.5$ Gyr. However, in this case no
cold filaments are developed. In conclusion, a very small change in
the flattening can compensate for a variation of the characteristic
scale length, showing again the strong sensitivity of models to
variations of galactic structural parameters in the CDPR scenario.

\section {Discussion}

Although the number of models we ran is not sufficient for an accurate
statistical and quantitative analysis of the gas flow properties in
flat and rotating galaxies, a qualitative comparison with the
spherical models discussed in CDPR and with cooling flow models is
possible; it is also interesting to compare our results with recent
observational findings (Pellegrini, Held \& Ciotti 1997).


In order to compare the behavior of S0 models with the equienergetical
spherical models discussed in CDPR, we performed calculations for
model galaxies with $\chi=1$ at the present epoch (models A1-A2).
We stress again
that while $\chi=1$ is a quite natural product of realistic spherical
models, it turns out to be rather extreme for flat models. In any case, in
flat models a clear X-ray underluminosity with respect to their
spherical counterpart is found, mainly due to a drastic reduction of
the hot gas mass. The amount of galactic rotation contributes only
slightly to such a reduction.


We also ran a few models with $\vtsn=0$ (models B1-B2).  These have a
remarkably low $\lx$, in contrast with spherical cooling flows which
in general show excessively high X-ray luminosities. Although the model
without rotation displays a present day $\lx$ similar to that of the
rotating model, it was much more luminous in the past. As a
consequence, with different choices of the structural and dynamical
parameters, a large difference in the $\lx$ can be obtained.  Such a
possibility is supported also by the models of BM, where different
degrees of rotation lead to a spread in $\lx$.


When the galaxy structure and dynamics are more similar to those of
observed galaxies, and SNIa's are present, $\chi$ is low, and the
galaxy is permanently in a global wind phase or loses a large amount
of gas during a Hubble time; the flow decoupling takes place at late
times, even for $\vtsn$ as low as 0.3.  The qualitative predictions
made in CP are thus correct: it is much easier to extract gas from a
flat, realistic galaxy model, than from a rounder system of the same
$\lb$.  $\lx$ may span all the values in the range between that of a
pure wind and approximately $\lsn$ by, for instance, varying $b$ or the
galaxy flattening, similarly to what happens for spherical models in
the CDPR scenario for small variations of the model parameters.

Moreover, the X-ray underluminosity of flat galaxies can naturally be
accounted for because their $\lx$ never reaches $\lsn$, even in the
inflow phase.  A possible explanation of the observations could be that at
low $\lb$, where the CDPR scenario predicts a significant number of
objects to have an energy budget sufficient to unbind the gas, flat
galaxies are preferentially found in the wind phase with respect to spherical
galaxies of the same optical luminosity, while at high $\lb$, where a
significant number of galaxies contain inflows, the underluminosity of
flat galaxies could be due to the underluminosity of their inflow
phase, as found (albeit for different reasons) in our models and by
BM.


We computed also a few models (not shown here) with a different flattening of
the dark matter distribution with respect to the
stellar component. The results are qualitatively similar to those
discussed above.  As a general rule, for a given stellar distribution,
a rounder halo increases $\chi$ and reduces $\tfil$. For example, a
model similar to A1 but with a spherical halo $(a/b)_{\rm h}=0$,
has $\chi=1.14$ and $\tfil\simeq\tcc = 1$ Gyr; a model similar to
C2 but with a rounder halo $(a/b)_{\rm h}=0.6$, has
$\chi=0.8$ and $\tfil\simeq 6.2$ Gyr. As expected, as the flattening
of the dark halo increases, $\chi$ decreases and $\tfil$ increases:
for example, a model similar to D2 but with $(a\b)_{\rm h}=1.3$
always remains in the wind phase, in accordance with the lower value
of $\chi =0.79$.


We conclude that, on the basis of $\lx$ alone, it is not possible to
decide whether the CDPR scenario or the cooling flow scenario is more
suitable to describe the flows in S0s and Es: both can reproduce the
X-ray underluminosity of flat galaxies.  The {\it scatter} in the
$\lx-\lb$ diagram (at a fixed optical luminosity) is mainly produced by
the galactic flattening (and rotation to a minor extent) in the CDPR
scenario, and by the amount of rotation in the cooling flow models.

In any case, the current explanations of the observational results are
unsatisfactory. In fact, Pellegrini et al. (1996) pose unsolved
problems, especially concerning the explanation of the scatter in the
$\lx -\lb$ diagram, as a function of the flattening and rotation. It is
found that $\lx /\lb$ does not correlate with rotation and/or
flattening. On the contrary, a very strong {\it segregation} effect is
found: only galaxies with small flattening and low $\vrot/\sigc$ show
a scatter in $\lx/\lb$, while flat objects and strongly rotating ones
are systematically found only at low $\lx/\lb$.  The main factor in
determining the $\lx/\lb$ segregation (and scatter, as a consequence)
could be related to the {\it slope} of the galaxy surface brightness
profile in the very center, as measured by HST observations
(Pellegrini 1997, and references therein). This would tell us that a
{\it nuclear} origin for the X-ray behavior of early-type galaxies
should be taken into proper account (Ciotti \& Ostriker 1997). It is
remarkable how the {\it independently chosen} separation of early-type
galaxies in cuspy and core galaxies, a separation operated using only
photometric criteria, reproduces exactly the segregation in the
$\lx-\lb$ plane: while core galaxies show the entire scatter in $\lx$
at fixed $\lb$, cuspy galaxies are invariably found only at low $\lx$.
 
Other problems affect both the CDPR models and the cooling flow
models.  For example, the absence of massive cold disks in Es seems to
support the CDPR scenario, while the claimed very low metallicity in
the hot gas is against it, because a low metallicity contrasts with
the assumption of a non negligible value of $\vtsn$.  In fact, in
principle, constraints on the SNIa rate can be given by estimates of
the iron abundance in galactic flows (see, e.g., \cite{rcdp93}, and
references therein). Under the assumption of solar abundance ratios,
the analysis of the available data suggests a very low iron abundance,
consistent with no SNIa's enrichment and even lower than that of the
stellar component (Ohashi et al. 1990; Awaki et al. 1991; Ikebe et
al. 1992; Serlemitsos et al. 1993; Loewenstein et al. 1994; Awaki et
al. 1994; Arimoto et al. 1997; Matsumoto et al. 1997).  This is rather
puzzling, in that a value as high as $\vtsn=0.3$ agrees with the
current optical estimates of the SNIa rates. However, some authors
have found that more complex multi-temperature models with higher
abundance give a better fit to the data (Kim et al. 1996; Buote \&
Fabian 1997).  So, there are arguments on both sides of the abundance
question, which is far from being a closed issue. In any case it is
clear that the CDPR scenario is ruled out in case a null SNIa rate is
clearly determined.


An interesting feature of our models, that could be used to
discriminate between the CDPR and the cooling flow scenario, is the
spontaneous occurrence of the flow decoupling.  This is at the origin
of the striking feature of the transient cold filaments, which -- if
caught during their formation -- should be observed first as soft
X-ray emitting structures, and subsequently as H$\alpha$ filaments. In
model A1 this takes place at $\tfil\simeq 2$ Gyr, while in model C1
(the model with $b=4$ kpc but similar to A1 in all other input parameters)
it occurs at $\tfil\simeq 10$ Gyr.  Due to this sensibility to the
galaxy structure, it is plausible that for a few galaxies the
decoupling of the flow takes place at the present epoch, and so these H$\alpha$
features could be observed\footnote{Note that a flow decoupling -- due
to different physical reasons -- is shown also in spherical power-law
galaxies with low dark matter content and low SNIa heating
(\cite{pc97}).}. Indeed, it is tempting to identify the transient cold
filaments with the soft X-ray clumps (\cite{kf95}) or H$\alpha$
filaments (\cite{tsa91}, Trinchieri, Noris, \& di Serego Alighieri
1997) recently observed.  We note how the size of the numerical and
observed structures (of the order of $\sim 10$ kpc) are surprisingly
similar. This could be an alternative explanation to the {\it galactic
drips} proposed by Mathews (1997).

A final comment is on the X-ray surface brightness distributions of
the computed models. As already found by BM for cooling flow models,
rotation is the key factor in determining whether $\SX$ is more or
less round than $\SS$, even in the presence of SNIa heating: rotation
produces a flatter $\SX$.  A favourable feature of the CDPR scenario
is that outflowing galaxies have rounder X-ray isophotes, which
alleviates the problem of excessively flat $\SX$ profiles, affecting
our cooling flow models as well as the models of KM and BM.  A
detailed observational analysis of the relationship between optical
and X-ray surface brightness for the E4 galaxy NGC 720 and S0 galaxy
NGC 1332 has been carried out by Buote \& Canizares (1997, and
references therein).  While data for NGC 1332 are more uncertain, an
analysis of the X-ray isophotes of NGC 720 shows that they are
misaligned with and rounder than the optical ones. The authors
conclude that the shape of $\SX$ is due to the flattening of the dark
halo, which is higher than that of the stellar component. In fact,
our simulations shows that, if rotation is the origin of flattening of
the X-ray isophotes, then a $decreasing$ flattening of $\SX$ with
galactocentric radius is expected, at variance with the observations.
On the other hand, in our simulations flatter dark halos produce
(during the inflow phase) flatter X-ray isophotes, as expected.

\section{Conclusions}

We have performed 2D numerical simulations of gas flows in S0 galaxy
models with different flattening, following the same recipe for the
input physics adopted in CDPR, and exploring the effects of different
amounts of ordered rotation and SNIa energy injection rates, and
structural parameters.  As a general rule, the WOI sequence which occurs
in the 1D models is replaced by a wind phase followed by a decoupled
inflow. The main results can be summarized as follows:

 
1. As expected models without SNIa heating ($\chi=4$) host a cooling
flow from the beginning. The flow is considerably ordered, and
essentially all the gas lost by the stars is cooled and accumulated in
the galactic center. If rotation is present, the cold material settles
in a disk on the galactic equatorial plane.

2. Models with SNIa heating and $\chi\simeq 1$ at the present epoch,
invert their
flow much earlier than the spherical ones with the same energy
budget. In particular, as the SNIa rate decreases, the flow tends to
revert its motion starting from the galactic polar regions. In this
phase of strong decoupling, cold filaments are created at the
interface between inflowing and outflowing gas. This strongly unstable
phase lasts $\sim 2$ Gyr, after which the filaments are definitively
accreted on the galactic center. Later on, the turbulences are unable to
create regions of very cool, dense gas. The temperature may however
fluctuate and gas cooler than X-ray temperatures is present. In the
rotating model, the presence of SNIa greatly reduces the formation of
the cold disk. In general, less cold matter is present in these models
($\simeq 10^{10}\msun$) than in those with $\chi =4$ ($\simeq
10^{11}\msun$).

3. Models with realistic values of the dynamical quantities are
characterized by low $\chi$ values.  While for model D1 ($b=5$ kpc,
$\chi =0.7$) the flow is always escaping without displaying any
interesting feature, in model C1 ($b=4$
kpc, $\chi =0.85$), the flow inversion occurs at $\tcc=5.5$ Gyr, and
filaments form at $\tfil\simeq 10.5$ Gyr. Of course, no cold mass is
present in the first model; in the second, $5\,10^9\msun$ of cold gas
form a small disk.

4. For any fixed value of $\vtsn$, rotating models are X-ray fainter
than the non-rotating ones, for the presence of the cold disk as well
as for a different distribution of the hot gas. When compared to 1D
models, 2D models are much fainter in the X-rays. The former flows, in fact,
have $\lx$ similar to or even higher than $\lsn$ after the flow
inversion; the latter, instead, always have $\lx<\lsn$. This is due to
several combined factors: the loss of a SNIa energy fraction which
is emitted by the gas below the X-ray temperature in high density
regions (which are much larger than in spherical models); the loss of
a SNIa energy fraction used by the gas still outflowing from the
external regions of the disk; the corresponding gravitational energy
not released by this escaping material, and, for rotating models, the
loss of thermalization of the gravitational energy of the cold
disk. Another main difference of the temporal evolution of $\lx$ is its
noisy behaviour in 2D models. This is due to the presence of
turbulence and to the formation of cold filaments.

5. We have shown that the flattening of the galaxy in the CDPR
scenario allows for a decoupling of the flow, in which different flow
regimes are present in different part of the galaxy at the same time.
We have found also that the flow decoupling is a general feature,
i.e., all our models, except those without SNIa heating and the model
with $\chi$ substantially lower than unity, show a flow decoupling.
We note however that the $\lx$ associated to a decoupled flow when $\chi
<1$, is not high: it is true, on the contrary, that {\it the flows are
decoupled even when $\chi >1$}, as the galaxy loses gas from the disk
even with a low SNIa rate.  So, the energetical analysis carried out
by CP is shown to be qualitatively verified: objects with a low $\chi$
value are found preferentially at low $\lx$, and at variance with the
cooling flow models the effect of the flattening is shown to be
important in determining the flow evolution.

6. For non-rotating, inflow models, the edge-on $\SX$ is rounder than
the corresponding $\SS$ of the parent galaxy; when rotation is
present, $\SX$ becomes flatter, and very similar to $\SS$. For models
with SNIa heating, the $\SX$ morphology evolves with time, from a
smooth, peanut-shaped geometry to a generally rounder shape than the
stellar profile, through disturbed phases during the presence of cold
filaments.  When rotation is present, in the galagtic center the $\SX$
isophotes are somewhat flatter than $\SS$, at variance with
non-rotating models in which $\SX$ remains more spherical up to the
center. As a general rule, even for low line-of-sight inclination
angles over the galactic equatorial plane, the isophotes of $\SX$ and
$\SS$ tend to become similar.

All these new features considerably enrich the old CDPR scenario, and
represent a step further in the direction of a more realistic
simulation of a situation that is proving to be very complex.  In
future, we plan to perform numerical simulations of gas flows in
models tailored on observed X-ray emitting S0 galaxies, in order to
improve our understanding of the much richer phenomenology of
aspherical gas flows.

\acknowledgments

We would like to thank James Binney, Fabrizio Brighenti, Silvia
Pellegrini, and Ginevra Trinchieri for useful discussions, and the
referee for comments that improved the paper. We are indebted to
Giovanna Stirpe who carefully read the manuscript. We would also like to
thank the CINECA Computing Center for having kindly provided the CRAY
C90 for our numerical computations.  This work was partially supported
by the Italian MURST and the Italian Space Agency (ASI) trough
grant ASI-95-RS-152.

\clearpage

\clearpage

\figcaption[]{The gas density and velocity fields of model A1 at four 
              different times (0.46, 2.16, 3.37, 15) Gyr (top left, top right, 
              bottom left, bottom right). In the top left panel the dashed
              lines represent the stellar density distribution. In the second 
              panel the large high density transient structure is apparent. 
              Note how the velocity field in the second and third panels is 
              dominated by large vortices, while in the fourth panel the 
              velocity is very low, and nearly random.
              \label{fig1}}

\figcaption[]{The gas temperature and velocity for model A1 (high 
              resolution simulation) at $t\simeq 2$ Gyr, when the cold 
              filament is completely developed. Note how the disk is still 
              strongly degassing.
              \label{fig2}}

\figcaption[]{The time evolution of the mass and luminosity of gas flows
              for the models discussed in the text. Panel (a) illustrates 
              the behaviour of the hot gas of model B1 ($\vtsn=0, k=1$; dashed
              line), and of model B2 ($\vtsn=0, k=0$; dotted line). The solid 
              line represents the total mass lost by the stars. Panel (c) 
              represents the $\lx$ evolution for the same models. Panel (b) 
              shows the mass budget for the models with $\vtsn=0.3$. 
              Dotted lines represent the evolution of the hot gas mass for 
              model A1 ($k=1$, line 1); A2 ($k=0$, line 2); C1 ($k=1$, line 3);
              D1 ($k=1$, line 4). Dashed lines show the total gas mass inside 
              the galaxy. The solid line again represents the total mass lost 
              by the stars. Panel (d) shows the evolution of $\lx$ for these 
              models (dotted lines), and the solid line represents $\lsn$.
              \label{fig3}}

\figcaption[]{The optical (solid lines) and X-ray (dotted lines) surface 
              brightness distributions for model A1, at the same   
              representative times as in Fig. 1 (time increases from top to 
              bottom). The optical isophotes are separated by one magnitude. 
              The X-ray isophotes are logarithmically spaced: the lowest value 
              (-8.51 in cgs units) is the same in all panels, the maximum is 
              -3.34 (a), -2.97 (b), -3.37 (c), -4.47 (d).
              \label{fig4}}

\figcaption[]{The gas temperature and velocity for model C1 at $t\simeq 10$ 
              Gyr, when the cold filament is completely developed.
              \label{fig5}}

\clearpage

\begin{deluxetable}{crrrrrrrrrrr}
\footnotesize
\tablecaption{Models parameters. \label{tbl-1}}
\tablewidth{0pt}
\tablehead{ \colhead{Model\tablenotemark{a}}  &  
            \colhead{$b$\tablenotemark{b}}    & 
            \colhead{$a/b$}  & 
            \colhead{$\sigma_{\rm c}$\tablenotemark{c}}  & 
            \colhead{$v_{\rm rot}$\tablenotemark{d}}  & 
            \colhead{$\vtsn$} & 
            \colhead{$k$} & 
            \colhead{$\chi$\tablenotemark{e}}  &
            \colhead{$t_{\rm cc}$\tablenotemark{f}} &  
            \colhead{$t_{\rm fil}$\tablenotemark{f}}  
          } 
\startdata

A1 & $3$ & $1$   & $330$ & $360$ & $0.3$  & $1$ & $1$    & $2$    & $2$    \nl

A2 & $3$ & $1$   & $330$ & $0$   & $0.3$  & $0$ & $1$    & $1.7$  & $1.7$  \nl

A3 & $3$ & $0.7$ & $376$ & $365$ & $0.3$  & $1$ & $1.20$ & $1.4$  & $1.4$  \nl

A4 & $3$ & $1.3$ & $300$ & $347$ & $0.3$  & $1$ & $0.94$ & $3$    & $3$    \nl
\nl
B1 & $3$ & $1$   & $330$ & $360$ & $0.0$  & $1$ & $4$    & $0$    & $...$  \nl
 
B2 & $3$ & $1$   & $330$ & $0$   &  $0.0$ & $0$ & $4$    & $0$    & $...$  \nl
\nl
C1 & $4$ & $1$   & $290$ & $309$ &  $0.3$ & $1$ & $0.85$ & $5.5$  & $10.5$ \nl

C2 & $4$ & $1.3$ & $260$ & $300$ &  $0.3$ & $1$ & $0.75$ & $...$  & $...$  \nl

D1 & $5$ & $1$   & $258$ & $277$ &  $0.3$ & $1$ & $0.70$ & $...$  & $...$  \nl

D2 & $5$ & $0.6$ & $305$ & $283$ &  $0.3$ & $1$ & $0.86$ & $5.5$  & $...$  \nl
 
\enddata

 
\tablenotetext{a}{All models have $\lb=10^{10}\lsun$, $\mas=10^{11}\msun$,
                  and $\mah/\mas=2$}
\tablenotetext{b}{in kpc}
\tablenotetext{c}{three-dimensional central velocity dispersion in km s$^{-1}$}
\tablenotetext{d}{maximum rotational velocity in km s$^{-1}$}
\tablenotetext{e}{for the rotating models computed under the assumption of
                  complete thermalization of ordered motions}
\tablenotetext{f}{in Gyr}
\end{deluxetable}

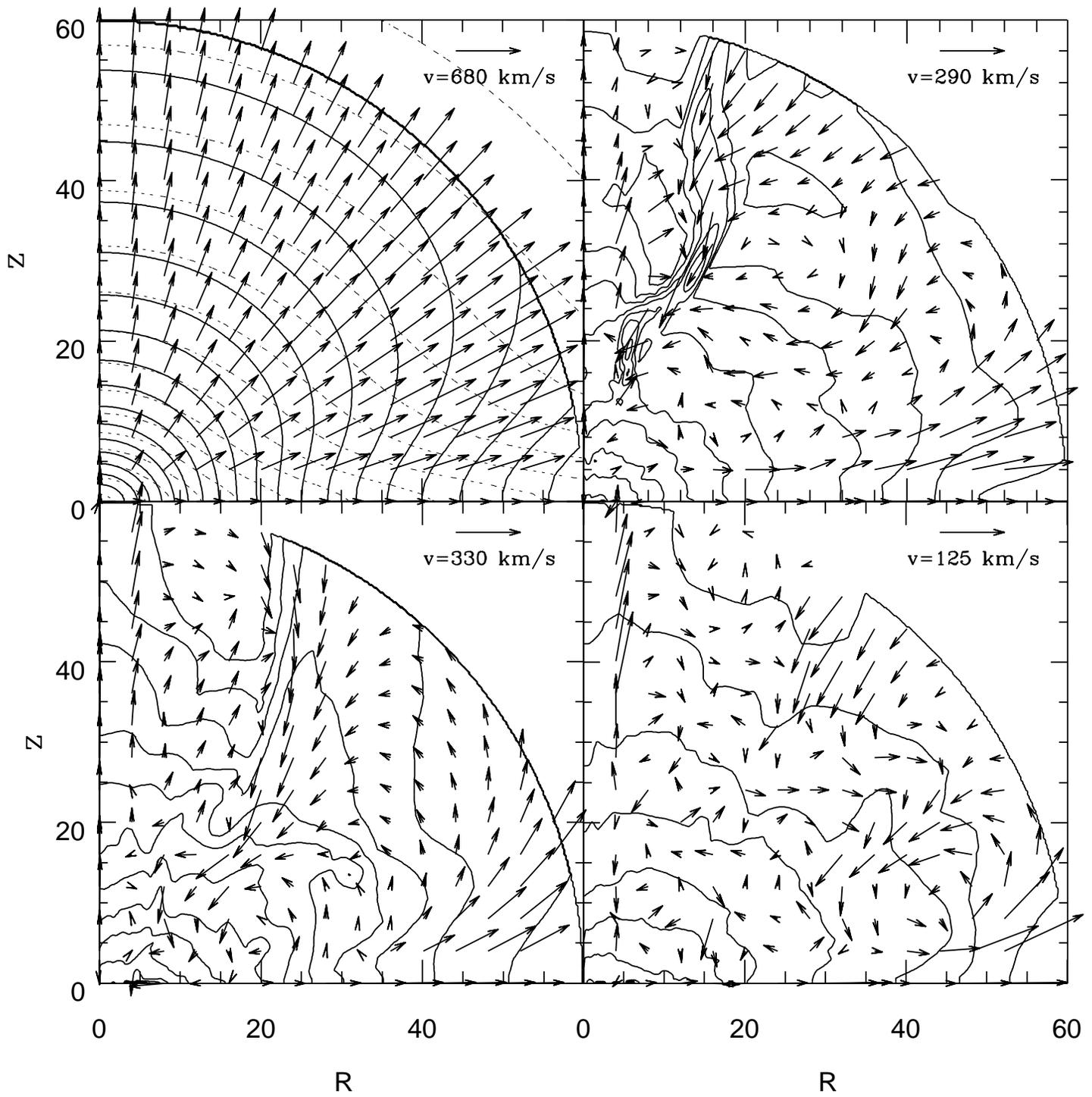
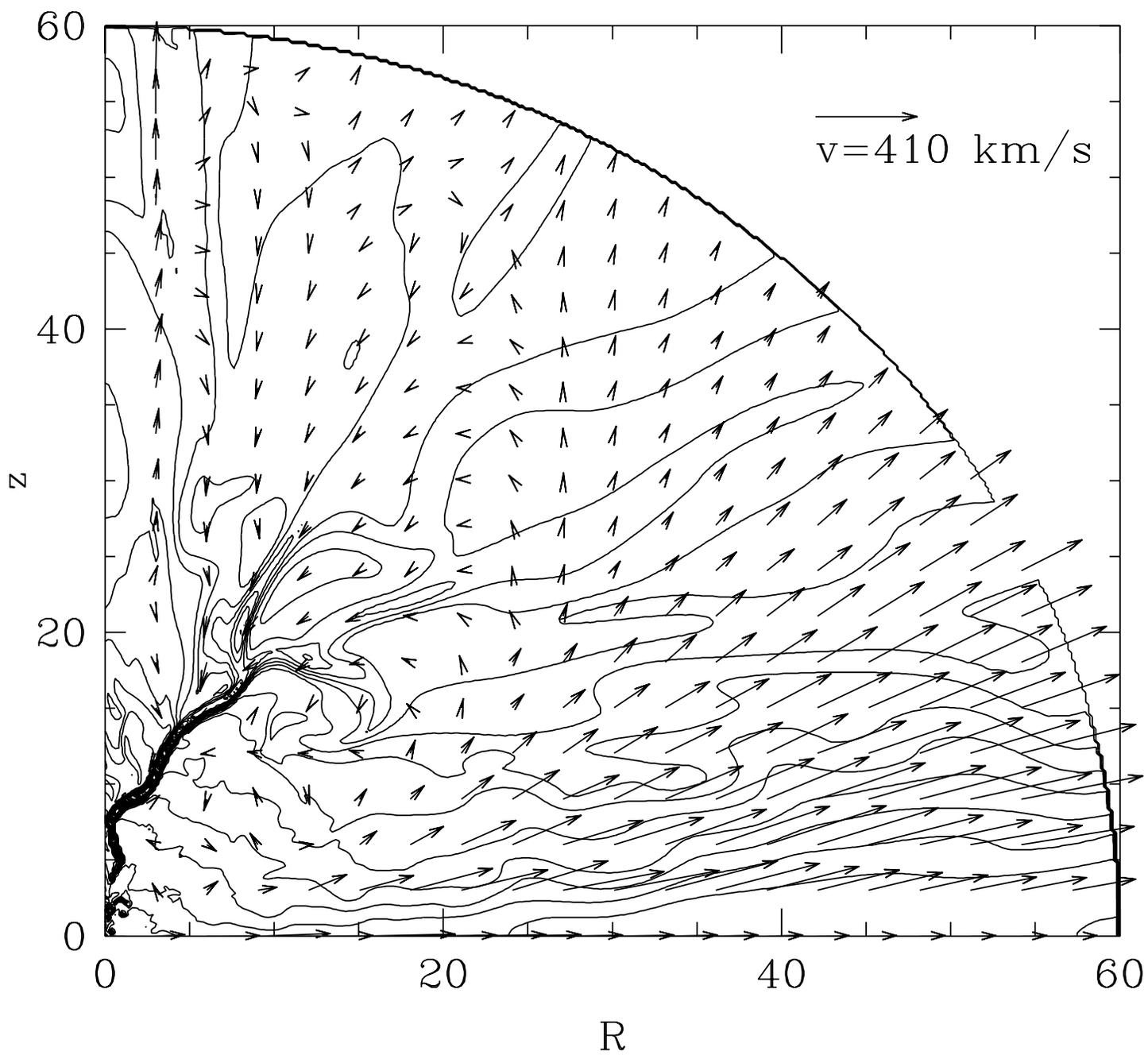
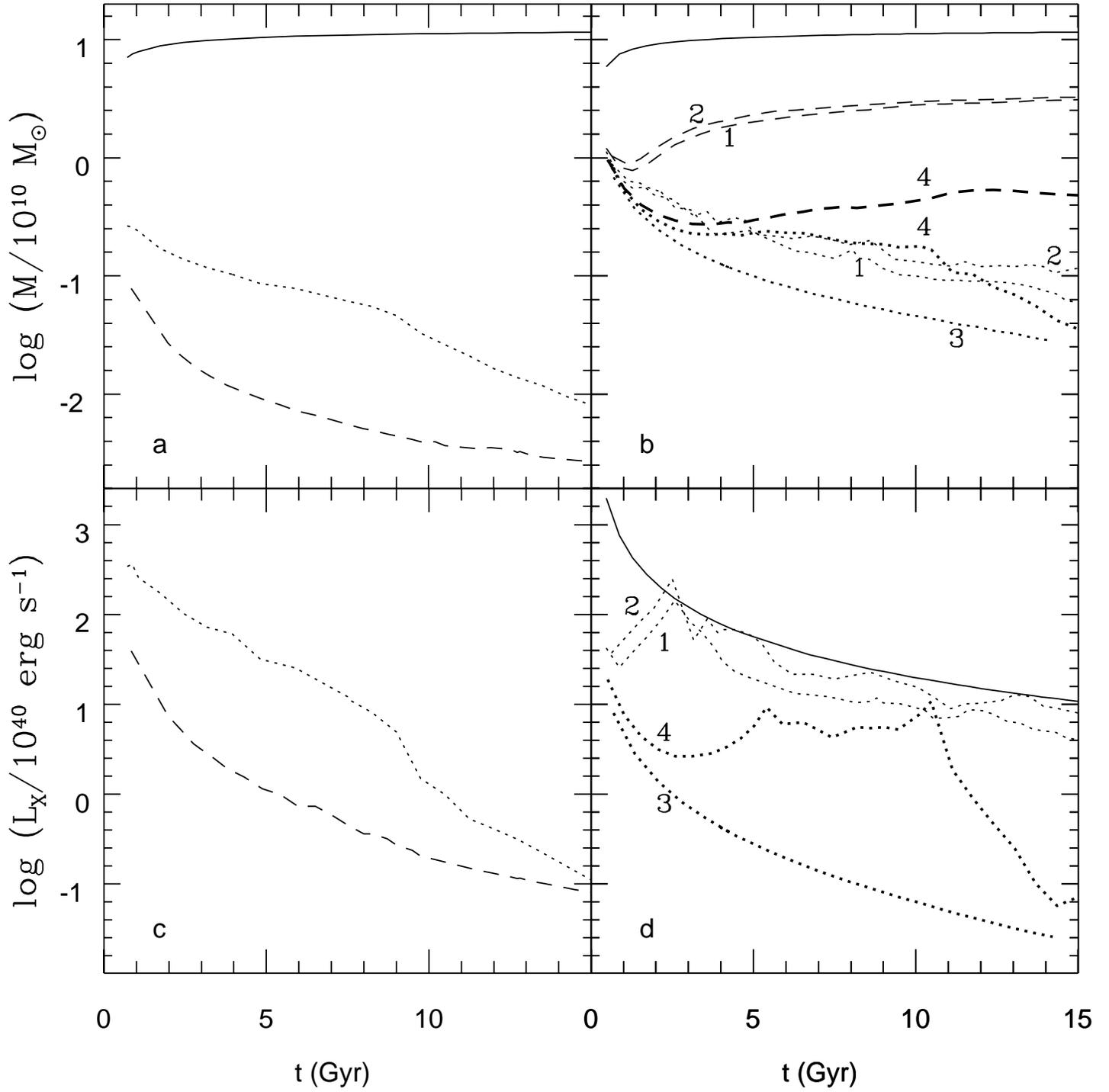
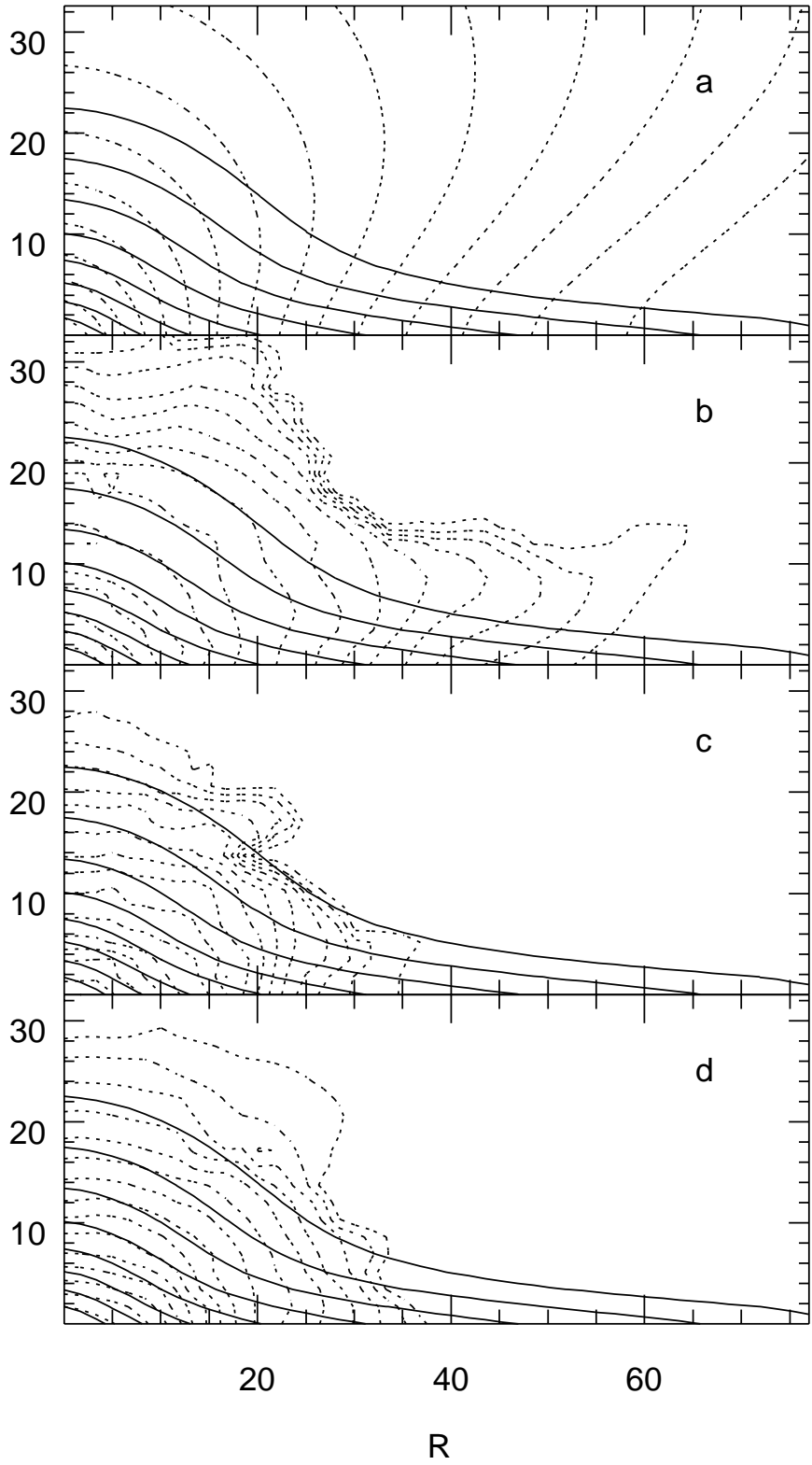
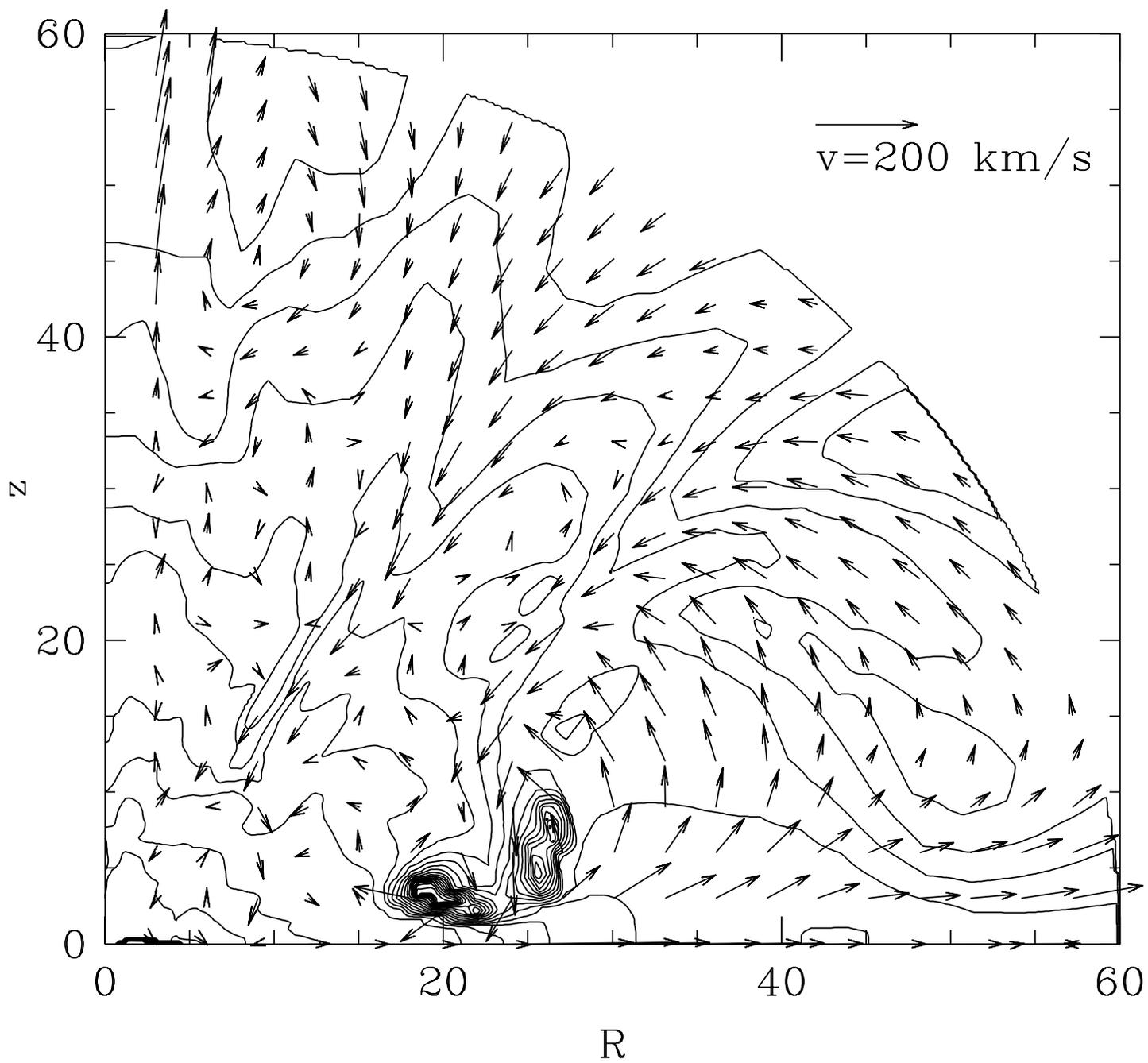
\end{document}